# Top Quark Mass and Supersymmetry

Stefan Pokorski[a][*][†][‡]

[a]Max–Planck Institute for Physics, Werner Heisenberg Institute,
Foeringer Ring 6, 80805 Munich, Germany

We summarize the expectations for the top quark mass in supersymmetric models and discuss the potential implications of its value measured by the CDF group in Fermilab.

## 1. Introduction

Evidence for the top quark has been recently reported by the CDF group at Fermilab, with the mass $M_t = (174 \pm 17)$ GeV. It is, therefore, interesting to summarize the expectations for the top quark mass in supersymmetric models and to discuss the potential implications of the measured value.

## 2. Top quark mass and precision tests of the minimal supersymmetric standard model (MSSM)

The MSSM can be viewed as the low energy effective theory of a wide class of supersymmetric models with R parity conservation [1]. It is important to test the model independently of its high energy origin as far as possible. Several groups (Altarelli et al., Chankowski et al., Ellis et al., Haber et al., Langacker et al.) proceed with systematic precision tests of the MSSM based on the electroweak data. This programme is not yet fully completed as a) not all process dependent one–loop supersymmetric corrections have been included, b) the parameter space of the MSSM has not been fully explored in the simultaneous analysis of all available electroweak data. The complete one–loop MSSM analysis exists, so far, only for the corrections to the mass of the W boson [2]: $M_W = f(G_\mu, \alpha_{EM}, M_Z, M_t, M_h, \ldots)$

[*]On leave from Institute for Theoretical Physics, Warsaw University, Hoza 69, 00–681 Warsaw, Poland.
[†]Supported by the Ministry for Science, Research and Culture of Land Brandenburg, contract II.1-3141-2/8(94).
[‡]Partially supported by a grant of the Polish Committee for Scientific Reseach.

which are conveniently parametrized by the quantity $\Delta r$:

$$G_\mu / \sqrt{2} = \frac{\pi \alpha_{EM}}{\left(1 - \frac{M_W^2}{M_Z^2}\right) M_W^2} [1+ \\ + \Delta r (\alpha_{EM}, M_W, M_Z, M_t, M_h, \ldots)] \qquad (1)$$

The dots stand for the additional parameters present in the MSSM. In the standard model, the measured value of $\Delta r = 0.044 \pm 0.015$ gives an upper bound on the top quark mass as a function of the Higgs boson mass $M_h$. In the MSSM, in spite of several additional free parameters, the following "theorem" holds [2]: for the same mass $M_h$ of the standard model Higgs boson and of the lighter Higgs boson in the MSSM, the standard model bound on $M_t$ obtained from $\Delta r$ cannot be softened and is reached only for soft slepton masses $M_{\tilde{l}} \geq 0(100$ GeV$)$ and soft squark masses $m_{\tilde{q}} \geq 0(200$ GeV$)$. Remembering that in the MSSM there exists the upper bound $M_h \leq 150$ GeV (for $M_t < 250$ GeV and $m_{\tilde{q}} < 2$TeV), one obtains from $\Delta r$ at $1\sigma$ level

$$M_t \leq 190 \text{ GEV} \qquad (2)$$

One should, however, be aware of the fact that the measured values of the other electroweak observables (in particular $\Gamma_{Z \to b\bar{b}}$) seem to constrain the top quark mass in the standard model from above much stronger than $\Delta r$. There, the MSSM may help to reconcile the heavy top quark with precision data [3] and it is clear that the extention of the type of analysis performed in ref. [2] to other observables is most welcome. It is also clear that, with the measured value of the top quark mass so close to the bound (2), such an analysis will put strong constraints on the sparticle masses. *Since the value of the top quark mass is measured, the precision electroweak data can be used to constrain the parameters of the MSSM.*

Another interesting issue is the significance of the measured value of the top quark mass for the minimal supersymmetric model with unification of strong and electroweak forces. This will be discussed in the rest of this lecture, by gradually supplementing the MSSM with additional assumptions.

## 3. Top quark mass and perturbative Yukawa couplings up to $0(10^{15} - 10^{19}$ GeV).

Present unification scenarios rely on perturbative physics. The requirement of the top quark Yukawa coupling to remain perturbative up to the GUT scale puts the upper bound on its value at $M_Z$:

$$h_t(M_Z) < h_t^{MAX}(M_Z) \equiv h_t^{IR} \quad (3)$$

where $h_t^{IR}$ is the so-called quasi-infrared fixed point value [4] of $h_t$. In the MSSM, assuming its validity all the way up to the GUT scale, one obtains then the following upper bound for the top quark running mass [5]:

$$M_t^{MAX} = \frac{h_t^{MAX}}{g_2} M_W \sin\beta$$
$$\cong (190 - 210) \text{ GeV } \sin\beta \quad (4)$$

where $\tan\beta = v_2/v_1$ is the ratio of the two Higgs vacuum expectation values and the range of the predicted values corresponds to the uncertainty in $\alpha_s$: $\alpha_s(M_Z) = 0.11 - 0.13$. The prediction for $M_t^{MAX}$(pole) (after inclusion of the two-loop QCD corrections to the running mass) as a function of $\tan\beta$ is plotted in Fig. 1.

The comment worth making at this point is that the measured $M_t$ is stunningly close to its quasi-infrared fixed point values, for generic values of $\tan\beta$ ! Is there any theoretical reason to expect that ?

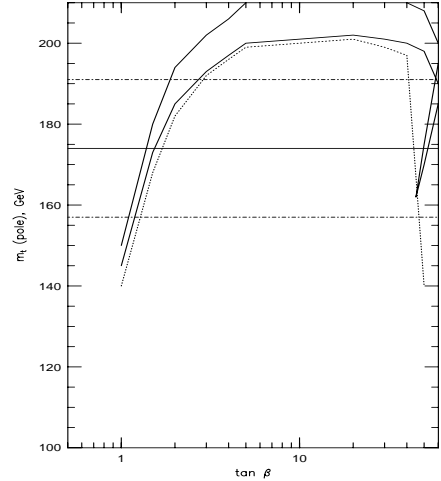

Figure 1. The top quark mass as a function of $\tan\beta$. The region bounded by solid lines: the quasi-infrared fixed point prediction with $\alpha_s = 0.11 - 0.13$. Dashed curve: the lower bound predicted by the b–$\tau$ Yukawa unification (the upper bound coincides with the IR upper bound). Narrow strip: the prediction following from the t–b–$\tau$ Yukawa unification. Horizontal lines: experimental result $M_t = (174 \pm 17)$ GeV.

## 4. Top quark mass and coupling unification in the MSSM

The gauge coupling unification is almost insensitive to the value of the top quark mass. However, $M_t$ is an important parameter for the bottom-tau Yukawa unification, which is predicted by simple grand unification scenarios (SU(5), SO(10)). This is due to the fact that the top quark Yukawa coupling, if large enough, affects the renormalization of the bottom Yukawa coupling in its running from $M_Z$ to the GUT scale according to the equation

$$4\pi \frac{d}{dt}\left(\frac{h_b}{h_\tau}\right) = \frac{h_b}{h_\tau}\left(\frac{16\alpha_s}{3} - 3\frac{h_b^2}{4\pi} - \frac{h_t^2}{4\pi} + 3\frac{h_\tau^2}{4\pi}\right) (5)$$

with $t = 2\ln\left(\frac{M_Z}{Q}\right)$.

It turns out that the strong interaction renormalization effects present in eq.(5) are, with $\alpha_s(M_Z) = 0.11 - 0.13$, too strong. Starting with the ratio $h_b/h_\tau = 1$ at the GUT scale and neglecting the Yukawa couplings in eq.(5) one obtains the bottom quark pole mass well above the experimental range $M_b = (4.9 \pm 0.3)$ GeV (a recent analysis of the B decays based on the QCD sum rules suggests even smaller value of $M_b$; R. Rueckl, private communication). Thus, indeed, a large top quark Yukawa coupling *is necessary* for the bottom-tau Yukawa unification in the MSSM.

An important implication of the measured value of the top quark mass is that, in the MSSM, in addition to the gauge couplings, also the b and $\tau$ Yukawa couplings unify at least within (20-30)% accuracy for a broad range of $\tan\beta$ values.

To make the relation between $M_t$ and b-$\tau$ Yukawa unification more precise it is convenient to reverse the problem: assume b-$\tau$ Yukawa unification and predict $M_t$ as a function of $\tan\beta$ (using $M_\tau$ and $M_b = (4.9 \pm 0.3)$ GeV as input parameters). This prediction is also shown in Fig. 1 and it almost coincides with the perturbative upper bound $M_t^{MAX}(\tan\beta)$ [6]. *The b-$\tau$ Yukawa unification implies the top quark mass to be very close to its quasi-infrared fixed point values.* Several comments to Fig. 1 are in order. First, the region in Fig. 1 bounded from below by the dotted line is the prediction which follows from the b-$\tau$ Yukawa *and* gauge coupling unification. For a detailed discussion of the role played by the gauge coupling unification to get the results shown in Fig. 1 we refer the reader to ref.[7]. Another remark is that relaxing the b-$\tau$ Yukawa unification by 10% gives the same predictions for $M_t$, with $M_b$(pole) =4.9 GeV, as the exact unification with $M_b = 5.2$ GeV. For larger deviations from $h_b = h_\tau$, the predicted values of $M_t$ fall below the IR values, e.g. for $h_b = 0.8 h_\tau$ the plateau is at $\sim$ 180 GeV. Finally, the IR fixed point values of $M_t$ and the predictions following from the b-$\tau$ Yukawa unification overlap in almost the whole region of $\tan\beta$, but with exception of large $\tan\beta$ region. There, the b-$\tau$ Yukawa unification does not imply such close proximity to the IR fixed point because $h_b$ renormalizes itself (we reach the region $h_b > h_t$). In the large $\tan\beta$ region the full unification $h_t = h_b = h_\tau$ is possible. If we impose it then the predicted value of $M_t$ is again closer to the IR fixed point values (the narrow strip in Fig. 1).

We can conclude as follows: The b-$\tau$ Yukawa coupling unification (and $Y_t = Y_b = Y_\tau$ at very large $\tan\beta$ values) implies $Y_t(M_Z)$ to be very close to its quasi-IR fixed point value. The value of the top quark mass is then strongly correlated with $\tan\beta$ and

a) $M_t$ within its present central experimental values $\approx$ (170-180) GeV (or below) corresponds to $\tan\beta < 2$ or $\tan\beta \cong M_t/M_b$ (see Fig. 1);

b) $M_t$ in the upper range of the presently reported values can even be consistent with $\tan\beta$ in the plateau region of Fig. 1.

The ($M_t$,$\tan\beta$) correlation which follows from the proximity of the $Y_t$ to the IR fixed point has important consequences for the prediction for the lightest Higgs boson mass $M_h$ in the MSSM. As is well known, the upper bound on the mass $M_h$ is the function of $\tan\beta$ and, therefore becomes now much stronger than the absolute upper bound for a given $M_t$. This is shown in Fig. 2 and is an encouraging message for the Higgs boson search at LEP2. (I am grateful to R. Barbieri for suggesting this plot to me.)

Finally, even as low a value of $M_t$ as 160 GeV is consistent with all values of $\tan\beta$ if we do not require the b-$\tau$ Yukawa unification to better than (20-30)% accuracy.

## 5. Top quark mass and radiative electroweak breaking

The top quark mass is an important parameter for the mechanism of radiative electroweak breaking. In the unification scenario the soft supersymmetry breaking parameters in the scalar potential are postulated to be (almost) universal at the GUT scale and radiative electroweak breaking is the necessary requirement for such models. The discussion in the previous section suggests two systematic ways of investigating the impact of the large top quark mass on predictions of the MSSM with radiative electroweak breaking:

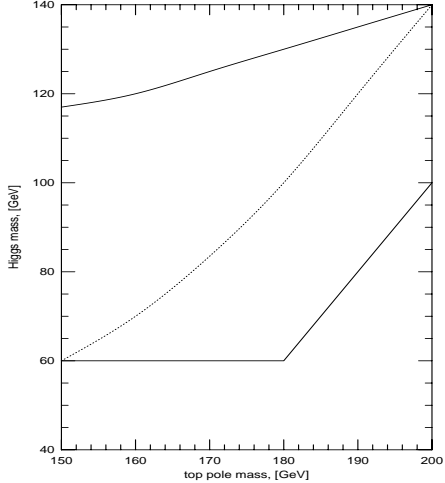

Figure 2. The upper bounds on the lightest Higgs boson mass as a function of the top quark mass. Solid upper curve: the absolute upper bound in the MSSM with $m_{\tilde{q}} < 2\text{TeV}$. Dotted line: the upper bound with the top quark Yukawa coupling close to its IR fixed point. Solid lower curve: the lower bound in the model with radiative elecrtoweak breaking and $Y_t$ close to its IR fixed point value.

(a) study the predictions of the model with $h_t \approx h_t^{IR}$ as a function of $M_t$ (i.e. as a function of $\tan\beta$) for, say, 160 GeV$< M_t <$200 GeV;

(b) study the predictions with, say, $M_t = 180$ GeV as a function of $\tan\beta$, for small and intermediate values of $\tan\beta$ (i.e. gradually departing from the $h_t \approx h_t^{IR}$ case)

(The large $\tan\beta$ case, $\tan\beta \approx M_t/M_b$ has several characteristic features which have to be discussed separately [8]).

In case (b), the parameter space consistent with radiative breaking is weakly dependent on $\tan\beta$. The two effects: increase of $\tan\beta$ and a gradual departure of $h_t(M_Z)$ from its quasi-infrared fixed point value partially cancel each other. Therefore, for the generic predictions of the model (with radiative electroweak breaking and the heavy top quark) for the superpartner mass spectrum it is sufficient to consider the case of $h_t \approx h_t^{IR}$.

The main characteristic pattern of the parameter space consistent with radiative electroweak breaking for $h_t \approx h_t^{IR}$ is the correlation between the Higgs mixing parameter $\mu$ and the universal gaugino and scalar masses at the GUT scale $M_{1/2}$ and $m_o$, respectively [9]:

$$\mu^2 + \frac{M_Z^2}{2} = m_o^2 \frac{1 + 0.5\tan^2\beta}{\tan^2\beta - 1}$$
$$+ M_{1/2}^2 \frac{0.5 + 3.5\tan^2\beta}{\tan^2\beta - 1} \quad (6)$$

The relation (6) follows from the requirement that the Higgs potential with quantum corrections included by the renormalization group evolution from the GUT scale has the proper minimum at the electroweak scale. This condition reads

$$\tan^2\beta = \frac{m_{H_1}^2 + \mu^2 + M_Z^2/2}{m_{H_2}^2 + \mu^2 + M_Z^2/2} \quad (7)$$

and the running of the Higgs mass parameters gives

$$m_{H_1}^2(M_Z) = m_o^2 + 0.5 M_{1/2}^2$$
$$m_{H_2}^2(M_Z) = -0.5 m_o^2 - 3.5 M_{1/2}^2 \quad (8)$$

The relation (6) is obtained from eqs.(7) and (8). The above discussion is based on the one-loop RGE which give the correct qualitative insight into the numerical results of ref.[9], obtained with two-loop running and the leading threshold effects included [10].

It follows from eq.(6) that

$$\mu > M_{1/2} \quad (9)$$

and, therefore, the lightest neutralino is strongly dominated by the gaugino component [11] (the low energy gaugino mass $M_2 \approx 0.8 M_{1/2}$, so that $\mu > M_2$). In consequence, the neutralino annihilation proceeds mainly through the slepton exchange (annihilation into quarks mediated by

squark exchange is strongly suppressed: $m_{\tilde{q}}^2 = 0.5m_o^2 + 6M_{1/2}^2 \gg m_\chi^2$):

$$\sigma_{ann} \sim \frac{m_\chi^2}{(m_{\tilde{l}}^2 + m_\chi^2)^2} \qquad (10)$$

where $m_\chi$ and $m_{\tilde{l}}$ are the lightest neutralino and slepton masses, respectively. The requirement that the neutralino relic abundance satisfies the condition $\Omega < 1$ ($\Omega \sim \sigma_{ann}^{-1}$) puts then strong upper bounds on the values of the neutralino and slepton masses. The bound on $m_\chi$ is stronger than the absolute bound $\mathcal{O}(1\ \text{TeV})$ of ref.[?] mainly because only leptonic annihilation channels are now effectively open. It translates itself in an obvious way ($m_\chi \sim 0.4 M_{1/2}$ and $M_2 \sim 0.8 M_{1/2}$) into an upper bound on $M_2$ and in consequence on the chargino mass. The bound on slepton masses can be relaxed only when $2m_\chi \approx M_h$ or $2m_\chi \approx M_Z$. The weak couplings of a gaugino–like neutralino to $h$ and $Z$ bosons (vanishing in the limit of pure gaugino) are balanced by the resonance and an acceptable relic abundance can be obtained, no matter how heavy the sleptons are (calculation of the annihilation cross section on a resonance needs special care [13]). Those predictions for the chargino and slepton masses are shown in Fig.3 and are generic for the model with radiative electroweak breaking *in the presence of the heavy top quark*, if acceptable relic abundance of the lightest neutralino is required.

The results for larger values of $\tan\beta$ are qualitatively similar, with the upper bound on the chargino mass slowly increasing with $\tan\beta$.

I am grateful to P. Chankowski, M. Olechowski, M. Carena, J. Rosiek, C. Wagner, A. Dabelstein, W. Hollik, P. Gondolo and W. Moesle for fruitful collaboration and to W. Moesle for his patient help in preparing the figures.

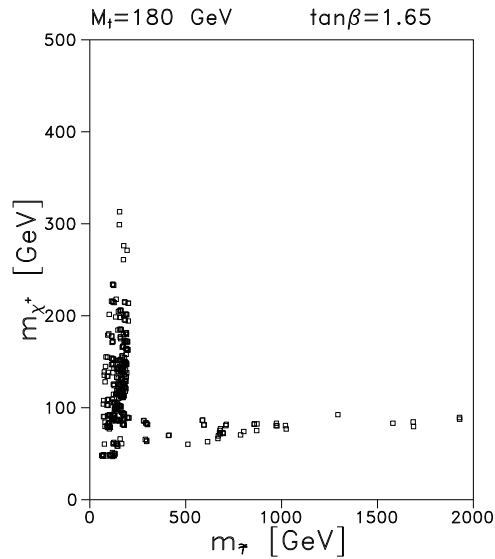

Figure 3. The region of the chargino and slepton masses consistent with $\Omega h^2 < 0.7$ in the model with radiative electroweak breaking.